\documentclass[cameraready]{Interspeech}

\title{MSR-HuBERT: Self-supervised Pre-training for Adaptation to Multiple Sampling Rates}

\author[affiliation={1}]{Zikang}{Huang}
\author[affiliation={1}]{Meng}{Ge}
\author[affiliation={1}]{Tianrui}{Wang}
\author[affiliation={1}]{Xuanchen}{Li}
\author[affiliation={1}]{Xiaobao}{Wang}
\author[affiliation={1,2},correspondingauthor=true]{Longbiao}{Wang}
\author[affiliation={3}]{Jianwu}{Dang}

\address{
  $^1$Tianjin Key Laboratory of Cognitive Computing and Application, Tianjin University, China \\
  $^2$Huiyan Technology Company, Ltd., China \\
  $^3$Chinese Academy of Sciences, China
}

\email{huangzikang@tju.edu.cn}

\keywords{self-supervised learning, sampling-rate adaptation, speech reconstruction, speech recognition, fine-tuning}

\usepackage{comment}
\usepackage{multirow}

\begin{document}
\maketitle
\begin{abstract}
    Self-supervised learning (SSL) has advanced speech processing. However, existing speech SSL methods typically assume a single sampling rate and struggle with mixed-rate data due to temporal resolution mismatch. To address this limitation, we propose MSRHuBERT, a multi-sampling-rate adaptive pre-training method. Building on HuBERT, we replace its single-rate downsampling CNN with a multi-sampling-rate adaptive downsampling CNN that maps raw waveforms from different sampling rates to a shared temporal resolution without resampling. This design enables unified mixed-rate pre-training and fine-tuning. In experiments spanning 16 to 48 kHz, MSRHuBERT outperforms HuBERT on speech recognition and full-band speech reconstruction, preserving high-frequency detail while modeling low-frequency semantic structure. Moreover, MSRHuBERT retains HuBERT's mask-prediction objective and Transformer encoder, so existing analyses and improvements that were developed for HuBERT can apply directly.
\end{abstract}

\section{Introduction}
Self-supervised learning (SSL) has enabled major advances in natural language processing and computer vision \cite{devlin2019bert,jing2020self}. In speech, SSL has emerged as a leading approach for learning powerful speech representations by pre-training on large unlabeled corpora using objectives that derive supervision from the signal itself rather than from human annotations \cite{mohamed2022self,wang2023adapter,lin2024selective,shi2023multi}. After pre-training, these models act as representation extractors and are fine‑tuned on labeled data for downstream tasks such as automatic speech recognition (ASR) \cite{zhao2022improving,huang2022investigating,radford2023robust}.

\begin{figure}[t]
\centering
\includegraphics[width=7.5cm]{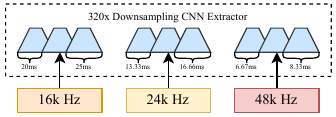}
\caption{
The resolution mismatch across diverse sampling rates with a fixed 320× downsampling CNN.
} 
\label{intro_pic}
\end{figure}

Mainstream speech SSL models (e.g., wav2vec 2.0, HuBERT, and WavLM) adopt a common design: a convolutional encoder compresses the raw waveform into frame‑level features with a fixed 20 ms frame shift, which serve as the basic units for pre-training and are realized by a 320× temporal downsampling for 16 kHz audio \cite{baevski2020wav2vec,hsu2021hubert,chen2022wavlm}. Consequently, these models are effectively tied to 16 kHz. For other sampling rates, the extracted frame-level features' frame shift and temporal resolution no longer match the expected 20 ms. This resolution mismatch leads the models to fail to function correctly during both pre-training and fine-tuning. We refer to this issue as the resolution mismatch problem, as illustrated in Fig.~\ref{intro_pic}. The problem prevents direct use of non‑16 kHz audio in pre-training, reduces data diversity, and limits applicability to downstream tasks that operate at other sampling rates \cite{ardila2019common,li2024masksr,schroter2022deepfilternet}. Addressing sampling‑rate compatibility is therefore crucial for robust and widely usable speech SSL. To our knowledge, this work represents a pioneering effort to explicitly explore solutions to this resolution mismatch in speech SSL.

Simple remedies are unsatisfactory: training separate models per sampling rate is costly, and resampling high‑rate audio to 16 kHz discards high‑frequency information that is useful for downstream performance \cite{abrol2024sampling}. Frequency‑domain techniques (e.g., subband decomposition or fixed‑duration STFTs) can achieve sampling‑rate independence in speech enhancement \cite{yu2023efficient,paulus2022sampling}, but they do not integrate naturally into existing SSL pipelines without disrupting the original paradigm, because most speech SSL models operate on the time-domain, and the few studies on frequency-domain adaptation target improving training efficiency rather than performance \cite{yang2023fast}.

In this paper, we propose MSRHuBERT, a multi-sampling-rate adaptive pre-training method. By introducing a multi-sampling-rate adaptive downsampling convolutional feature extractor, routing audio of different sampling rates through a convolutional encoder with rate-specific downsampling, it maps inputs to frame-level features on a common temporal resolution, thereby alleviating the resolution mismatch problem. Meanwhile, we retain the standard SSL paradigm, the training objective and architecture. We implement our design on a HuBERT backbone and conduct pre-training and downstream fine-tuning on datasets sampled at multiple rates. Results show our approach achieves comparable performance on the ASR task that depends on low‑frequency signal content, while improving performance on the full‑band speech reconstruction task that relies on high‑frequency information. Moreover, because we retain the original SSL objective and architecture, improvements developed for HuBERT can be incorporated into our method to yield further gains. Finally, as the first work to explicitly mitigate multi‑sampling‑rate resolution mismatch in speech SSL, we empirically quantify the severe degradation that mismatch induces on ASR and speech reconstruction. Our main contributions are: (1) MSRHuBERT, a speech SSL framework adaptive for multiple sampling rates, including a multi‑sampling‑rate adaptive downsampling convolutional extractor that avoids resolution mismatch while preserving the core SSL paradigm; (2) empirical evidence across multiple sampling rates that the learned representations by our method retain low‑frequency content that is important for ASR while also capturing high‑frequency details beneficial for reconstruction; (3) the first identification and formalization of the resolution mismatch problem in speech SSL.
\section{HuBERT}
HuBERT \cite{hsu2021hubert} is a typical speech SSL framework. It benefits from an offline clustering (e.g., k-means) to generate frame-level pseudo labels $U = [u_1, u_2, ..., u_T]$, where $T$ is the number of speech frames. A fixed 320x downsampling convolutional neural network (CNN) $F(\text{·})$ converts 16 kHz speech $s_{16\text{k}}$ into the frame-level feature $H = [h_1, h_2, ..., h_T]$ with a frame shift of 20 ms, which is then fed to a Transformer encoder $G(\text{·})$ to get the contextual representation $C = [c_1, c_2, ..., c_T]$. During pre-training, the frame-level features $H$ are masked randomly before they are passed to the Transformer encoder. The whole pipeline can be formalized as
\begin{equation}
    C = [c_1, c_2, ..., c_T] = G(M(F(s_{16\text{k}})))\text{,}
    \label{eqn:hubert1}
\end{equation}
where $M(\text{·})$ is the mask operation on frame-level features. Finally, the model is trained to predict the pseudo labels of the masked frames using a cross-entropy (CE) loss function
\begin{equation}
    \mathcal{L}_{CE}(C, U) = - \sum_{t\in O}\log  p(u_t|c_t)\text{,}
    \label{eqn:hubert2}
\end{equation}
where $O$ denotes the set of indices masked in $H$. Note that, when applied to downstream tasks, $H$ is no longer masked and the obtained unmasked contextual representation $C$ is used.
\section{Proposed Method}
\label{sec:msrhubert}
\begin{figure}[t]
\centering
\includegraphics[width=7.5cm]{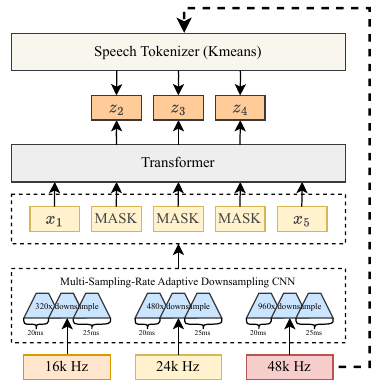}
\caption{
The architecture of MSRHuBERT, which takes raw waveforms at different sampling rates as input. The multi-sampling-rate adaptive downsampling CNN can map waveforms to a common temporal resolution by designing rate-specific downsampling, and supports mixed-rate pre-training.
} 
\label{architecture_pic}
\end{figure}

Fig.~\ref{architecture_pic} illustrates the overall architecture of MSRHuBERT. MSRHuBERT is a speech SSL framework that processes speech sampled at multiple sampling rates and is explicitly designed to resolve the resolution mismatch problem. Given speeches at diverse sampling rates, the model maps them to a common temporal resolution and performs mixed‑rate pre-training using a shared codebook. MSRHuBERT departs from HuBERT in a principal respect: a multi‑sampling‑rate adaptive downsampling CNN, which aligns frame‑level features across sampling rates, enabling direct pre-training on raw multi‑rate waveforms while preserving the intended 20 ms frame shift, modeling both low- and high-frequency information.

\subsection{Multi‑sampling‑rate Adaptive Downsampling CNN}
\label{ssec:msrcnn1}
The multi‑sampling‑rate adaptive downsampling CNN is designed to compress speech signals recorded at different sampling rates into frame‑level features that share a common temporal resolution without resampling, thereby tackling the resolution mismatch problem while preserving high‑frequency information. Specifically, given an input waveform $s_{a\text{k}}$ sampled at $a$ kHz, the model routes it to a rate-specific downsampling CNN $F_{dr}(\text{·})$ with downsampling ratio $dr$ so that the resulting frame‑level features match HuBERT's temporal convention of 20 ms frame shift. Inspired by \cite{yuslimmable}, we append a layer normalization after each downsampling CNN, which removes feature aggregation inconsistency across rate-specific CNNs and maps all extracted features into a shared feature space by normalizing mean and variance independently per branch \cite{ba2016layer}.
\begin{equation}
H = [h_1, h_2, \ldots, h_T] = \mathrm{LN}_{dr}\big(F_{dr}(s_{a\text{k}})\big),
\end{equation}
where $\mathrm{LN}_{dr}$ denotes layer normalizing attached to $F_{dr}$, and downsampling ratio $dr = a\text{k} \times 0.02$, achieved through carefully designing the stride and kernel width of each $F_{dr}$ (detailed in Table~\ref{tbl:design}). This design ensures that all sampling rates are aligned to the same temporal resolution before being fed to the shared Transformer encoder. 

Compared to HuBERT, the adaptive downsampling CNN enables direct processing of raw high‑sampling-rate speech without resampling, thereby preserving high‑frequency information, increasing the diversity of pre-training data, and extending applicability to downstream tasks operating at multiple sampling rates. Because the extracted features all share the standard temporal scale and lie in a unified feature space, the following mask prediction and Transformer encoder can be retained.

\subsection{Mixed-rate Mask Prediction via Single Codebook}
\label{ssec:msrcnn2}
Because the multi‑sampling‑rate adaptive downsampling CNN maps inputs from different sampling rates to frame‑level features on a common temporal grid. This unified temporal scale enables the subsequent use of a shared Transformer encoder and a single shared codebook for consistent contextual modeling. This unified formulation preserves the original training paradigm while allowing the model to exploit the greater diversity of multi‑rate pre-training data. Furthermore, this design ensures training efficiency, minimizes additional model parameters, and facilitates the effective extraction of robust speech representations.
Concretely, mask prediction is performed on features with a 20 ms frame shift, a convention used in prior work to yield effective speech representations. Following the HuBERT paradigm, the frame-level features, produced by the adaptive downsampling branch, are masked randomly and passed to a shared Transformer encoder whose contextual outputs are trained to predict offline clustering pseudo labels, as
\begin{equation}
\mathcal{L} = -\frac{1}{T}\sum_{t\in O}\log\frac{\text{exp}(\text{sim}(Ac_t, e_u)/\tau)}{ {\textstyle \sum_{u^\prime}^{U}\text{exp}(\text{sim}(Ac_t, e_{u^\prime})/\tau)} }   
\end{equation}
where $A$ is a projection matrix, $e_u$ is the embedding for pseudo label $u$ of frame $t$, $T$ is the number of masked frames, and $\tau$ scales the logit which is set to 0.1. 

This mixed‑rate single shared codebook training enables the model to absorb low‑ and high‑frequency cues present across different sampling rates, and constrain extracted contextual representations from diverse rates to lie in a common feature space, yielding rate‑invariant representations that simplify fine‑tuning on mixed‑rate downstream tasks.

\section{Experiments}
\label{sec:experiments}
\subsection{Pre-training Setup}
\label{ssec:pretraining_setup}
For multi‑sampling‑rate pre-training, we retain the 960‑hour LibriSpeech corpus \cite{panayotov2015librispeech} at 16 kHz used by HuBERT Base to ensure a fair comparison. For the other sampling rates we use the clean subset of the DNS Challenge 2022 dataset \cite{dubey2024icassp}, originally recorded at 48 kHz. This subset is uniformly resampled to 22.05 kHz, 24 kHz and 48 kHz, producing roughly 193 hours of speech for each rate.

All pre-training experiments are implemented using the Fairseq toolkit \cite{ott2019fairseq}. We train all models from random initialization \cite{lin2023melhubert} for 400k steps on four 24GB NVIDIA 4090D GPUs, using a batch size of 87.5 seconds of audio per GPU and eight times gradient accumulation to match HuBERT's effective batch configuration. 
During pre-training, we control same sampling-rate in each batch but mix batches from different sampling-rates within each update because of distributed training and gradient accumulation, allowing efficient learning from diverse data with minimal extra forward computation. Regarding parameters, our proposed CNN branches contribute little to the parameter count, which is mainly from the Transformer Encoder. Quantitatively, adding one sampling rate increases parameters by only 3\%, and the time cost of mixed training with 4 rates is just 2.6\% longer than single-rate training.
Other hyperparameters for the model are kept the same as in \cite{hsu2021hubert}. We use the HuBERT base as the baseline and backbone model, and the architecture of our proposed multi‑sampling‑rate adaptive downsampling CNN is shown in Table~\ref{tbl:design}. Note that HuBERT Base baseline is pre-trained for one sampling rate using the corresponding single-rate CNN.

\subsection{Fine-tuning Setup}
\label{ssec:finetuning_setup}
To evaluate the generality of representations learned by MSRHuBERT across different sampling rates, we perform fine‑tuning experiments under the SUPERB evaluation protocol \cite{yang2021superb}. In this protocol, the pre-trained model is kept frozen and only the lightweight downstream module is learnable.

\begin{table}[t]
    \centering
    \fontsize{7.5}{7.5}\selectfont
    \def\arraystretch{2.0}
    \setlength{\tabcolsep}{7.5pt}
    \setlength{\abovetopsep}{1pt}
    \setlength\belowbottomsep{0pt} 
    \setlength\aboverulesep{0pt} 
    \setlength\belowrulesep{0pt}
\caption{Architecture design for proposed multi-sampling-rate adaptive downsampling CNN. For the CNN branch of each sampling rate, the stride and kernel width is determined based on the calculation formulas for the downsampling rate, stride, and kernel width, as well as the prime factor decomposition.}
\scalebox{1}{
\begin{tabular}{c|c|c}
\toprule
\textbf{Sampling Rate} & \textbf{Strides} & \textbf{Kernel width}\\
\midrule
16 kHz & 5, 2, 2, 2, 2, 2, 2 & 10, 3, 3, 3, 3, 2, 2\\
22.05 kHz & 7, 7, 3, 3& 19, 14, 4, 3 \\
24 kHz & 5, 3, 2, 2, 2, 2, 2& 10, 5, 3, 3, 3, 2, 2\\
48 kHz & 5, 3, 2, 2, 2, 2, 2, 2& 10, 5, 3, 3, 3, 3, 2, 2\\
\bottomrule
\end{tabular}}
\label{tbl:design}
\end{table}

\begin{table*}[t]
    \centering
    \fontsize{7.5}{7.5}\selectfont
    \def\arraystretch{2.2}
    \setlength{\tabcolsep}{2.1pt}
    \setlength{\abovetopsep}{1pt}
    \setlength\belowbottomsep{0pt} 
    \setlength\aboverulesep{0pt} 
    \setlength\belowrulesep{0pt}
\caption{The comprehensive results for ASR and SR downstream tasks using different pre-trained models.}
\scalebox{1}{
\begin{tabular}{c|cccc|cccc|cccc}
\toprule
\multirow{3}{*}{\textbf{Model}} & \multicolumn{8}{c|}{\textbf{ASR (WER↓)}} & \multicolumn{4}{c}{\textbf{Full-band SR (STOI↑)}} \\

\cline{2-13}

& \multirow{2}{*}{\textbf{16k Hz}}& \multirow{2}{*}{\textbf{22.05k Hz}} & \multirow{2}{*}{\textbf{24k Hz}} & \multirow{2}{*}{\textbf{48k Hz}} & \multicolumn{4}{c|}{\textbf{four sampling rates mixed fine-tuning}} & \multirow{2}{*}{\textbf{16k Hz}} & \multirow{2}{*}{\textbf{22.05k Hz}} & \multirow{2}{*}{\textbf{24k Hz}} & \multirow{2}{*}{\textbf{48k Hz}}\\

& & & & & \textbf{16 kHz} & \textbf{22.05 kHz} & \textbf{24 kHz} & \textbf{48 kHz} & & & & \\
\midrule 
HuBERT Base & 6.41 & 3.34 & 6.84 & 5.96 & 7.42 & 3.35 & 7.61 & 6.63 & 88.46 & 93.04 & 88.52 & 81.42 \\
\midrule
\multicolumn{1}{l|}{Re. 16kHz HuBERT Base} & 6.03 & \textbf{3.15} & 6.59 & \textbf{5.61} & 6.89 & 2.95 & 6.97 & \textbf{5.53} & 89.35 & 93.46 & 88.49 & 82.49 \\
- w/o resampling in fine-tuning & 6.03 & 4.47 & 15.61 & 38.18 & 7.69 & 3.72 & 10.30 & 33.72 & 89.35 & 91.84 & 86.71 & 75.53 \\
\multicolumn{1}{l|}{Re. 24kHz HuBERT Base} & 6.15 & 3.28 & 6.54 & 5.70 & 6.95 & 3.02 & 7.24 & 5.74 & 89.61 & 94.32 & \textbf{89.34} & 83.75 \\
\multicolumn{1}{l|}{Re. 48kHz HuBERT Base} & 6.37 & 3.41 & 6.72 & 5.95 & 7.28 & 3.29 & 7.45 & 6.11 & 89.75 & 94.08 & 89.07 & \textbf{85.93} \\
\midrule
MSRHuBERT& \textbf{5.89} & 3.35 & \textbf{6.35} & 5.83 & \textbf{6.54} & \textbf{2.90} & \textbf{6.82} & 5.56 & \textbf{90.26} & \textbf{94.38} & 89.25 & 85.79 \\
\bottomrule
\end{tabular}}
\label{tbl:main_result}
\end{table*} 
For each sampling rate, we select a corresponding dataset: train-clean-100 and test-clean subsets of LibriSpeech for 16 kHz \cite{panayotov2015librispeech}, LJSpeech for 22.05 kHz \cite{ljspeech17}, train-clean-100 and test-clean subsets of LibriTTS for 24 kHz \cite{zen2019libritts}, and VCTK for 48 kHz \cite{veaux2017cstr}. We evaluate two downstream tasks that emphasize complementary spectral information: automatic speech recognition (ASR), which primarily depends on low‑frequency content, and full‑band speech reconstruction (SR), which additionally requires high‑frequency detail. For the speech reconstruction task, we adapt the original HiFi‑GAN vocoder \cite{kong2020hifi}: the vocoder utilizes the representation produced by the frozen pre‑trained model and reconstructs the waveform \cite{tan2024system,defossez2022high}. To ensure a fair comparison across pre‑trained models, the downstream architectures and hyperparameters are identical for each evaluated model.

\subsection{Evaluation on SR and ASR}
\label{ssec:evaluation_SR}

To evaluate MSRHuBERT's applicability across sampling rates, we compare five pre-training configurations: 1) HuBERT Base: Pre-trained only on the 960‑hour LibriSpeech at 16 kHz; 2) Resampled 16k HuBERT Base: Pre-trained on the full multi‑rate pre-training datasets after resampling all speech to 16 kHz; 3) Resampled 24k HuBERT Base: Pre-trained on the full multi‑rate pre-training datasets after resampling all speech to 24 kHz; 4) Resampled 48k HuBERT Base: Pre-trained on the full multi‑rate pre-training datasets after resampling all speech to 48 kHz; 5) MSRHuBERT: Pre-trained on the full datasets without resampling using our proposed architecture. During downstream fine‑tuning and evaluation, we resample each downstream dataset to the sampling rate expected by the pre-trained model so that frame‑level temporal resolution aligns with the pre-training convention and the resolution mismatch is avoided. To directly quantify the cost of violating this assumption, we additionally conduct experiments in which the downstream dataset is not resampled.

Table~\ref{tbl:main_result} presents comprehensive results for ASR and SR downstream tasks using different pre-trained models. From these results we draw four main observations: 1) The performance of the HuBERT depends systematically on the sampling rate used during pre-training. Models pre-trained at 16 kHz implicitly focus on low‑frequency semantic cues and consequently achieve stronger ASR performance, whereas models pre-trained at 48 kHz retain high‑frequency detail and therefore perform better on full‑band speech reconstruction, but the additional high‑frequency information interferes with the model's ability to concentrate on low‑frequency semantic structure, degrading ASR.
2) Our MSRHuBERT, by design, directly adapts to different sampling rates without resampling. As a result, it preserves high‑frequency information that benefits reconstruction tasks while still maintaining effective modeling of low‑frequency semantic content required by ASR. 3) Because the ASR training objective is the sampling‑rate invariant transcription text, we also conduct a mixed-rate fine-tuning experiment,
and our method exhibits promising performance. This empirical result supports the feasibility of mixed‑rate training for tasks whose labels are independent of sampling rate. 4) When the resolution mismatch exists between the sampling rate used for pre-training and the rate used for fine‑tuning, the convention learned during pre-training is violated and performance degrades. In addition, degradation increases with the magnitude of the sampling rate discrepancy.

\subsection{Exploration for Learning Paradigm}
\label{ssec:paradim}
Architecturally, MSRHuBERT retains HuBERT's core learning paradigm: mask prediction objective and the Transformer encoder remain unchanged. Based on the essence of the problem, we address the resolution mismatch problem by replacing the fixed single‑rate CNN with a multi‑sampling‑rate adaptive downsampling CNN that operates directly on time‑domain waveforms. As a result, existing layer‑wise analyses \cite{pasad2023comparative} and improvements developed for HuBERT can be transferred to MSRHuBERT with minimal modification. We verify this viewpoint both qualitatively and quantitatively.

\begin{figure}[t]
\centering
\includegraphics[height=4.3cm, width=7.5cm]{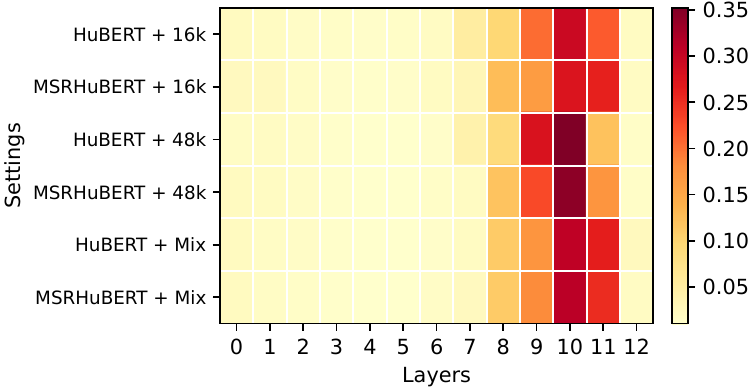}
\caption{
Weight analysis on the ASR task of the SUPERB Benchmark. Layer 0 corresponds to the input of the first Transformer layer. The y-axis represents different settings, including pre-trained model and sampling rate of the downstream dataset in ASR.} 
\label{weight}
\end{figure}

Following the SUPERB policies, we apply a weighted sum to the hidden states of different layers and feed it to the task-specific layers. Fig.~\ref{weight} shows the weights of different layers of HuBERT and MSRHuBERT models on the ASR task. The larger weight indicates the greater contribution of the corresponding layer. The two models exhibit qualitatively similar weight distributions, with dominant contributions arising from the same middle-to-upper layer region \cite{chen2022wavlm}. This correspondence indicates that MSRHuBERT sustains the layer-wise representational structure that HuBERT relies on.
\begin{table}[t]
    \centering
    \fontsize{7.5}{7.5}\selectfont
    \def\arraystretch{2.2}
    \setlength{\tabcolsep}{2.5pt}
    \setlength{\abovetopsep}{1pt}
    \setlength\belowbottomsep{0pt} 
    \setlength\aboverulesep{0pt} 
    \setlength\belowrulesep{0pt}
\caption{The result of MSRHuBERT with adaptation of intermediate layer supervision and progressive decoupling in downstream fine-tuning ASR task.} 
\scalebox{1}{
\begin{tabular}{c|cccc}
\toprule
\multirow{2}{*}{\textbf{Model}} & \multicolumn{4}{c}{\textbf{ASR (WER↓)}} \\
\cline{2-5}
& \textbf{16k Hz}& \textbf{22.05k Hz}& \textbf{24k Hz}& \textbf{48k Hz}\\
\midrule
\multicolumn{1}{l|}{Re. 16kHz HuBERT Base} & 6.03 & \textbf{3.15} & 6.59 & 5.61 \\
\multicolumn{1}{l|}{MSRHuBERT}& 5.89 & 3.35 & 6.35 & 5.83 \\
\multicolumn{1}{l|}{+ Intermediate Layer Supervision} & \textbf{5.56} & 3.17 & 6.16 & 5.67 \\
\multicolumn{1}{l|}{+ Progressive Decoupling} & 5.63 & 3.30 & \textbf{5.94} & \textbf{5.52} \\
\bottomrule
\end{tabular}}
\label{tbl:impro}
\end{table} 

To validate compatibility with existing methods for HuBERT, we introduce two representative enhancements, intermediate layer supervision \cite{wang2022improving} and progressive decoupling \cite{wang2025progressive}, developed for HuBERT to MSRHuBERT and evaluate their effects. Table~\ref{tbl:impro} reports the fine‑tuning ASR results: both improvements yield additional performance gains when applied to MSRHuBERT. These quantitative results confirm that MSRHuBERT retains HuBERT's training paradigm and therefore benefits from prior advances, demonstrating practical transferability and generalization across sampling rates. Moreover, our approach can be easily adapted to other sampling‑rate scenarios, such as 32 kHz or 44.1 kHz, with minimal modifications to the multi‑sampling‑rate adaptive downsampling CNN architecture. Specifically, we choose a downsampling factor matched to each input sampling rate so that all waveforms are compressed to a common temporal resolution prior to subsequent processing.
\section{Conclusion}
\label{sec:conclusion}
This paper proposes the resolution mismatch problem for speech SSL across diverse sampling rates and presents MSRHuBERT to mitigate this issue, using a self-supervised pre-training framework that introduces a multi-sampling-rate adaptive downsampling CNN without resampling and holds the core learning paradigm. A series of experiments have been carried out to show that our method preserves high-frequency detail and maintains effective modeling of low-frequency semantic content, sustaining the core learning paradigm and the layer-wise representational structure.


\bibliographystyle{IEEEtran}
\bibliography{mybib}

\end{document}